\documentclass[conference, 9pt]{IEEEtran}
\usepackage{hyperref}

\IEEEoverridecommandlockouts

\usepackage{fancyhdr}
\fancypagestyle{firstpageheader}{
    \fancyhf{}
    
    \fancyhead[C]{\Large \textbf{Accepted to the IEEE International Conference on Physical Assurance and Inspection of Electronics (PAINE) 2025}}
}

\usepackage{amsmath,amsfonts}
\usepackage{graphicx}
\usepackage{textcomp}
\usepackage{cite}
\usepackage{diagbox}
\usepackage{booktabs}
\usepackage{colortbl}
\usepackage{multirow}
\usepackage{tabularx}
\usepackage{hyperref}
\usepackage{subfig} 
\usepackage{wrapfig}
\usepackage{comment}
\usepackage{algorithm}
\usepackage[noend]{algpseudocode}
\algtext*{EndWhile}
\algtext*{EndIf}
\algtext*{EndFor}
\algnewcommand\algorithmicforeach{\textbf{for each}}
\algdef{S}[FOR]{ForEach}[1]{\algorithmicforeach\ #1\ \algorithmicdo}
\usepackage{xcolor}
\definecolor{myblue}{HTML}{93CDDD}
\definecolor{mygreen}{HTML}{17891B}
\definecolor{myyellow}{HTML}{DCC60F}
\definecolor{myorange}{HTML}{EB801B}
\usepackage[normalem]{ulem}
\usepackage{tikz}
\newcommand{\circled}[1]{%
  \tikz[baseline=(char.base)]{
    \node[draw, circle, fill=white, inner sep=1pt, line width=0.7pt] (char) {\scriptsize\bfseries #1};}}

\usepackage{enumitem}
\usepackage{footnote}
\usepackage{pifont}

\title{AuthenTree: A Scalable MPC-Based Distributed Trust Architecture for Chiplet-based Heterogeneous Systems}

\author{\IEEEauthorblockN{Ishraq Tashdid\textsuperscript{1}, Tasnuva Farheen\textsuperscript{2}, Sazadur Rahman\textsuperscript{1}}
\IEEEauthorblockA{\textsuperscript{1}University of Central Florida, Orlando, FL, USA, \textsuperscript{2}Louisiana State University, Baton Rouge, LA, USA\\
\{ishraq.tashdid, sazadur.rahman\}@ucf.edu\textsuperscript{1}, tfarheen@lsu.edu\textsuperscript{2}.}}

\begin{document}
\maketitle
\thispagestyle{firstpageheader} 

\begin{abstract}
The rapid adoption of chiplet-based heterogeneous integration is reshaping semiconductor design by enabling modular, scalable, and faster time-to-market solutions for AI and high-performance computing. However, multi-vendor assembly in post-fabrication environments fragments the supply chain and exposes SiP systems to serious security threats, including cloning, overproduction, and chiplet substitution. Existing authentication solutions depend on trusted integrators or centralized security anchors, which can expose sensitive data or create single points of failure. We introduce \emph{AuthenTree}, a distributed authentication framework that leverages multi-party computation (MPC) in a scalable tree-based architecture, removing the need for dedicated security hardware or centralized trust. \emph{AuthenTree} enables secure chiplet validation without revealing raw signatures, distributing trust across multiple integrator chiplets. Our evaluation in five SiP benchmarks demonstrates that \emph{AuthenTree} imposes minimal overhead, with an area as low as $0.48\%$ ($7{,}000~\mu\mathrm{m}^2$), an overhead power under $0.5\%$, and an authentication latency below $1~\mu\mathrm{s}$, bypassing previous work in some cases $700\times$. These results establish \emph{AuthenTree} as an efficient, robust, and scalable solution for next-generation chiplet-based security in zero-trust SiP environments.

\end{abstract}


\section{Introduction}
\label{sec:intro}

\begin{figure}[t]
    \centering
    \includegraphics[width=\linewidth]{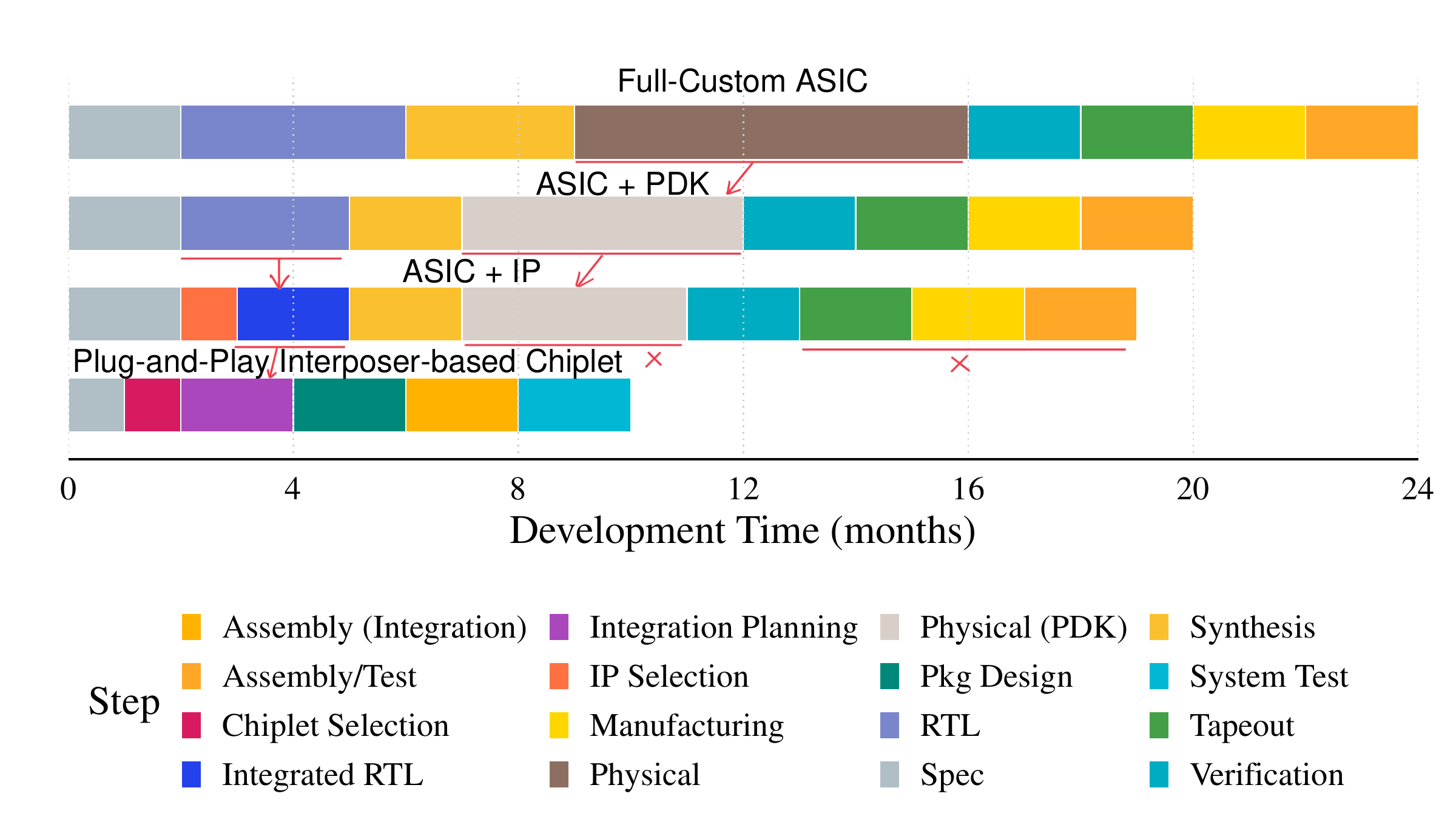}
    \caption{Comparative development timelines for custom ASIC, PDK-enabled, IP-enabled, and chiplet plug-and-play flows. Chiplet-based integration dramatically reduces time-to-market by enabling parallel development and assembly of pre-validated modules.}
    \label{fig:chiplet_ttm_motivation}
\end{figure}

\noindent
System-in-Package (SiP) platforms have rapidly emerged as a transformative paradigm in semiconductor integration, enabling the combination of diverse functionalities and process nodes through modular, pre-validated chiplets. This approach offers substantial advantages over traditional monolithic SoC design, including significantly enhanced design flexibility, dramatic reductions in non-recurring engineering (NRE) costs, and the ability to bring complex products to market on accelerated schedules~\cite{toshi, 2.5_3D_hetero, tashdid2025beyondppa, lau2019heterogeneous, tashdid2025ecologic, yahyaei2025ai}. By allowing designers to assemble systems from a portfolio of pre-fabricated chiplets—often sourced from multiple vendors or technology generations, SiPs facilitate true plug-and-play innovation, supporting rapid feature updates and customization with minimal fabrication overhead (see Fig.~\ref{fig:chiplet_ttm_motivation}). However, this open and collaborative design model introduces an array of new security and trust challenges. Specifically, the SiP assembly process is susceptible to adversarial actions such as the insertion of counterfeit chiplets, unauthorized replacement or overproduction, and post-fabrication tampering~\cite{gate_sip, ieee1838_casestudy}. As the supply chain grows more distributed and opaque, traditional methods, which typically depend on centralized trust anchors, dedicated security modules, or computationally intensive cryptographic primitives, become increasingly inadequate, introducing performance bottlenecks and substantial area, power, or cost overhead~\cite{pqc_hi, toshi}. Addressing these risks calls for the development of lightweight, distributed authentication frameworks that scale seamlessly with the number of chiplets, preserve strong system-wide trust guarantees, and integrate efficiently into modern heterogeneous design flows~\cite{lau2021chiplet}.


Despite recent advancements, existing security mechanisms for SiP-based heterogeneous integration remain limited in three critical aspects: centralized trust dependency, poor scalability, and unbalanced overhead-security tradeoffs. Many approaches retrofit SoC-era protections, such as logic locking~\cite{rahman2020defense}, watermarking~\cite{watermark_actiwate}, and IC metering~\cite{alkabani2007active}, into SiP designs without accounting for the expanded and untrusted supply chain, including interposer foundries, chiplet vendors, and SiP integrators. Others adopt dedicated security chiplets or hardware security modules~\cite{pqc_hi, sect_hi, gate_sip}, which inherently rely on a trusted chiplet, creating a \circled{1}~single point of trust. This design becomes a liability in modern SiPs comprising tens or hundreds of chiplets, where centralized verification quickly leads to \circled{2}~scalability bottlenecks. Moreover, these techniques often impose significant area and performance penalties, lacking a balanced tradeoff between strong security guarantees and \circled{3}~low implementation overhead, making them impractical for lightweight, composable SiP platforms.

\begin{table}[t]
\caption{Comparison with State-of-the-Art Authentication Solutions.}
\centering
\label{tab:intro_comparison}
\resizebox{0.48\textwidth}{!}{
\setlength{\tabcolsep}{1pt}
\begin{tabular}{ccc}
\toprule
\textbf{Technique} & \textbf{Limitations} & \textbf{Advantage in Our Scheme} \\
\midrule
\textbf{GATE-SiP~\cite{gate_sip}} &
\begin{tabular}[c]{@{}c@{}}TAP-based authentication;\\Vulnerable to MITM attacks\end{tabular} &
\begin{tabular}[c]{@{}c@{}}No TAP modification;\\Resilient to MITM attacks\end{tabular} \\
\midrule
\textbf{PQC-HI~\cite{pqc_hi}} &
\begin{tabular}[c]{@{}c@{}}High computational overhead;\\Susceptible to probing\end{tabular} &
\begin{tabular}[c]{@{}c@{}}Lightweight distributed\\authentication, strong\\signature obfuscation\end{tabular} \\
\midrule
\textbf{SECT-HI~\cite{sect_hi}} &
\begin{tabular}[c]{@{}c@{}}Limited to test encryption;\\Restricts vendor security control\end{tabular} &
\begin{tabular}[c]{@{}c@{}}Supports vendor and\\integrator security\end{tabular} \\

\midrule
\textbf{SAFE-SiP~\cite{tashdid2025safe}} &
\begin{tabular}[c]{@{}c@{}}Centralized integrator;\\Integrator compromise\\affects all authentication\end{tabular} &
\begin{tabular}[c]{@{}c@{}}Distributed trust;\\No single validator;\\Scalable, low-overhead\end{tabular} \\

\bottomrule
\end{tabular}}
\end{table}


Multi-party Computation (MPC) enables a set of mutually untrusted entities to jointly compute a function over their private inputs, guaranteeing both privacy and correctness. TinyGarble~\cite{tinygarble} and MPCircuits~\cite{mpcircuits} have shown that MPC protocols can be implemented efficiently in hardware, making secure computation a practical tool for system-level security. Recent efforts, such as Garblet~\cite{hashemi2025garblet} and Sipmeter~\cite{rahman2025sipmeter}, demonstrate the feasibility of using MPC techniques for secure computation and metering in chiplet-based platforms while maintaining low implementation overhead. SAFE-SiP~\cite{tashdid2025safe} extends this progress to authentication, proposing an MPC-based scheme for chiplet identity verification. However, \uline{all mentioned approaches in Tab.~\ref{tab:intro_comparison} still depend on a single trusted integrator chiplet for the final authentication, creating a centralized point of trust that, if compromised, threatens the security of the entire system.} These limitations prompt us to ask the following research questions:

\underline{\textbf{(RQ1)}} Can we architect an authentication framework for SiP systems that does not rely on any single chiplet as the root of trust? 

\underline{\textbf{(RQ2)}}~Is it possible to efficiently detect compromised or malicious integrator chiplets within a distributed authentication framework, while maintaining scalability for large SiP systems?

\underline{\textbf{(RQ3)}}~To what extent can such a distributed framework adapt to varying chiplet supply chain scenarios, including dynamic, plug-and-play chiplet addition or removal, without compromising authentication or system efficiency?

Inspired by advances in secure computation, we adopt distributed MPC protocols to enable robust, privacy-preserving chiplet authentication, even in the presence of untrusted supply chain entities. In this paper, we present \emph{AuthenTree}, an MPC-based authentication framework that decentralizes trust by leveraging collaborative verification among integrator chiplets, enabling secure and scalable authentication for heterogeneous SiP assemblies without exposing sensitive signature data. Our main contributions are:
\vspace{-0.05in}
\begin{enumerate}[leftmargin=*]
    \item We introduce \emph{AuthenTree}, a distributed authentication framework for heterogeneous SiP platforms, which leverages multi-party computation to decentralize trust across multiple integrator chiplets. \textbf{\uline{To the best of our knowledge, this is the first chiplet authentication architecture to realize scalable, zero-trust, plug-and-play security for SiP systems.}}
    \item We present a thorough security analysis, showing that \emph{AuthenTree} is resilient to a broad range of attacks, including compromise, chiplet forgery, replay, and denial-of-service, by detecting malicious behavior and preventing unauthorized authentication.
    \item Our hardware evaluation demonstrates that \emph{AuthenTree} imposes minimal overhead, with area impact as low as $0.48\%$ and authentication latency below $1~\mu\mathrm{s}$ across five diverse chiplet-based SiP benchmarks, while supporting $\sim0.56\%$ power overhead.
    \item To accelerate further research and facilitate industrial adoption, we will release our Verilog implementation, evaluation scripts, and benchmark data as open-source at our GitHub\footnote{\url{https://github.com/IshraqAtUCF/MPC_Based_Authentication}}.

\end{enumerate}

\vspace{-0.05in}

\noindent

The remainder of this paper is organized as follows: Section~\ref{sec:background} reviews the threat model and related work; Section~\ref{sec:method} details the \emph{AuthenTree} architecture and protocol; Section~\ref{sec:security-analysis} analyzes security; Section~\ref{sec:results} presents results; and Section~\ref{sec:conclusion} concludes the paper.

\begin{table}[t]
\centering
\caption{Attacks in SiP Supply Chain from Stakeholder Perspectives.}
\label{tab:attackvectors_perspective}
\setlength{\tabcolsep}{2pt}
\renewcommand{\arraystretch}{1.2}
\resizebox{0.48\textwidth}{!}{%
\begin{tabular}{ccc}
\toprule
\textbf{Perspective} & \textbf{Untrusted Entities} & \textbf{Primary Attack Vectors} \\
\midrule
Vendor & Integrator, Foundry & Chiplet IP theft, duplication \\
\midrule
Integrator & Vendors, Foundry & Counterfeit chiplets, malicious boot \\
\midrule
- & Foundry & Trojan insertion, Overproduction \\
\bottomrule
\end{tabular}
}
\end{table}

\section{Background}
\label{sec:background}

\noindent
This section introduces the zero trust paradigm in SiP design, presents the corresponding threat model, reviews limitations of prior approaches, and motivates the need for a distributed, scalable solution.

\subsection{Zero-Trust Paradigm} \label{subsec:zero_trust}

\noindent
In SiP platforms, the principle of zero trust dictates that every party in the supply chain, chip manufacturer, integrator, or foundry, must assume that other entities may be untrusted or adversarial. For vendors, this translates to protecting intellectual property and design information, preventing unwanted disclosure or exposure to side-channel attacks, while still demonstrating the authenticity of their chiplets. Integrators, who bear ultimate responsibility for system security, must assemble chiplets from diverse sources while guarding against the risk of malicious, counterfeit, or cloned components. Meanwhile, integrators strive to fully utilize available resources to create and validate complex heterogeneous systems. A genuine zero trust architecture must therefore rigorously safeguard the privacy and autonomy of all participants, while enabling reliable authentication across the entire assembly process. These zero trust considerations directly inform our subsequent threat model.

\subsection{SiP Supply Chain and Threat Model}\label{subsec:threat_model}

\noindent
Building on the zero trust paradigm, our threat model identifies practical risks at every stage of the SiP supply chain. Chiplet vendors can introduce counterfeit or malicious components; foundries and packaging facilities can insert hardware Trojans, enable unauthorized overproduction, or tamper with chiplet integration. Adversaries may exploit side-channel leakage or probing to extract sensitive data at any point, and even integrators, despite their system-level responsibilities, cannot be presumed fully trustworthy, as they could potentially clone, misuse, or manipulate authentication artifacts. No single entity is inherently trusted. Accordingly, an effective authentication framework must guarantee the inclusion of only genuine and untampered chiplets, maintain the confidentiality of all proprietary data, and robustly address risks such as cloning, substitution, and modification. 

\begin{figure}[t]
    \centering
    \includegraphics[width=\linewidth]{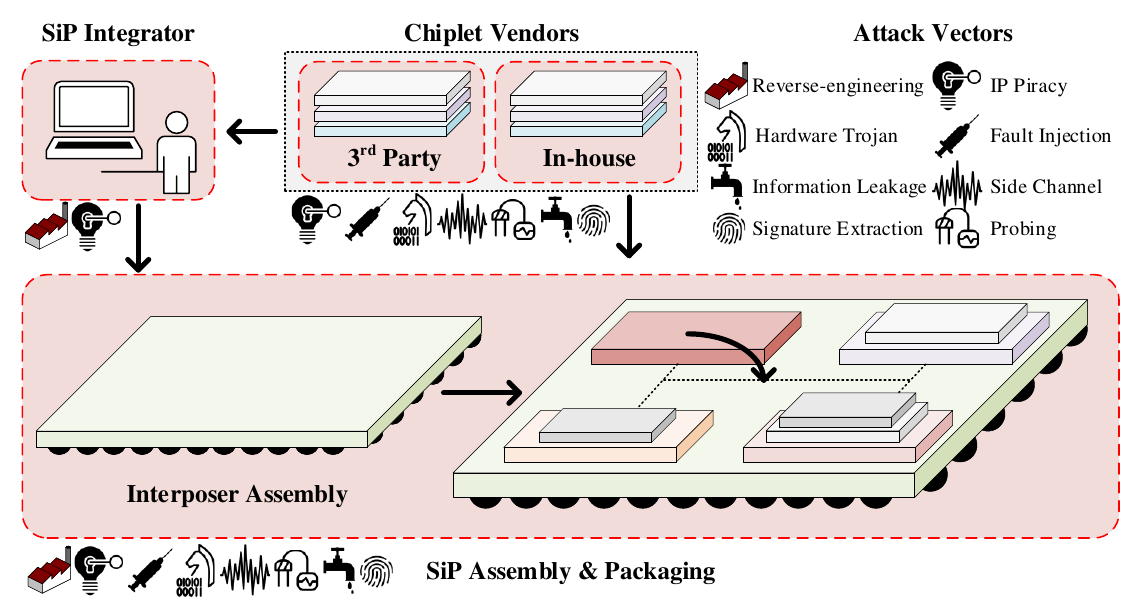}
    \caption{SiP design flow and life-cycle using third-party and in-house chiplets. Untrusted and trusted parties are marked by red and green boxes, respectively.}
    \label{fig:threatmodel}
\end{figure}

\subsection{Existing Works and their Drawbacks}\label{subsec:existing_techniques}

\noindent
Multiple lines of research have sought to mitigate these threats using fabrication-level defenses, hardware modules, or cryptographic protocols. Approaches such as split manufacturing (SM) and network-on-interconnect (NoI) obfuscation hide critical design information and complicate physical attacks~\cite{splitcore}, but SM’s reliance on trusted backend foundries restricts scalability and can introduce costly yield challenges, while NoI obfuscation complicates integration~\cite{postfab_lle}. Hardware module-based solutions, such as Chiplet Hardware Security Modules (CHSM) and Chiplet Security Intellectual Property (CSIP), offer authentication but at the cost of increased area, complexity and new single points of failure~\cite{toshi}. More recent cryptographic techniques (e.g., PQC-HI and SECT-HI) leverage post-quantum authentication and encrypted test protocols~\cite{pqc_hi, sect_hi}, but also add significant overhead and still rely on centralized trust anchors. Even advanced, MPC-based frameworks such as SAFE-SiP~\cite{tashdid2025safe} depend on a single integrator chiplet for final authentication, thus inheriting a bottleneck and trust vulnerability. Taken together, these limitations highlight the need for efficient, scalable, and truly distributed authentication.

\subsection{Motivation}\label{subsec:motivation}

\noindent
The persistent limitations of existing SiP security mechanisms, most notably centralized trust anchors, poor scalability, and excessive overhead, underscore the need for a fundamentally new approach. Many current solutions either retrofit protections from SoC-era designs without accounting for the expanded, adversarial SiP supply chain, or rely on dedicated hardware modules that quickly become bottlenecks and attack targets as system complexity grows. As a result, these methods struggle to meet the demands of lightweight, modular, and dynamically evolving SiP platforms, often incurring prohibitive area and performance costs. 

Motivated by these challenges and in direct response to our three central research questions, we develop a three-pronged strategy:
\begin{enumerate}[leftmargin=1.5em]
    \item \textbf{\textsc{Distributed Trust:}} Enable collaborative authentication among integrator chiplets, eliminating single points of failure and directly addressing \underline{\textbf{(RQ1)}}.
    \item \textbf{\textsc{Scalable Security:}} Employ a tree-based, hierarchical multi-party computation framework designed to efficiently scale with increasing chiplet count and diversity, supporting \underline{\textbf{(RQ2)}}.
    \item \textbf{\textsc{Dynamic Adaptability:}} Ensure modularity and plug-and-play operation so the system can securely adapt to chiplet addition or removal, a necessity for \underline{\textbf{(RQ3)}}.
\end{enumerate}

These three prongs form the basis for \emph{AuthenTree}; a scalable, distributed authentication architecture to address the unique security, scalability, and flexibility demands of next-generation SiP platforms.

\begin{figure}[t]
    \centering
    \includegraphics[width=\linewidth]{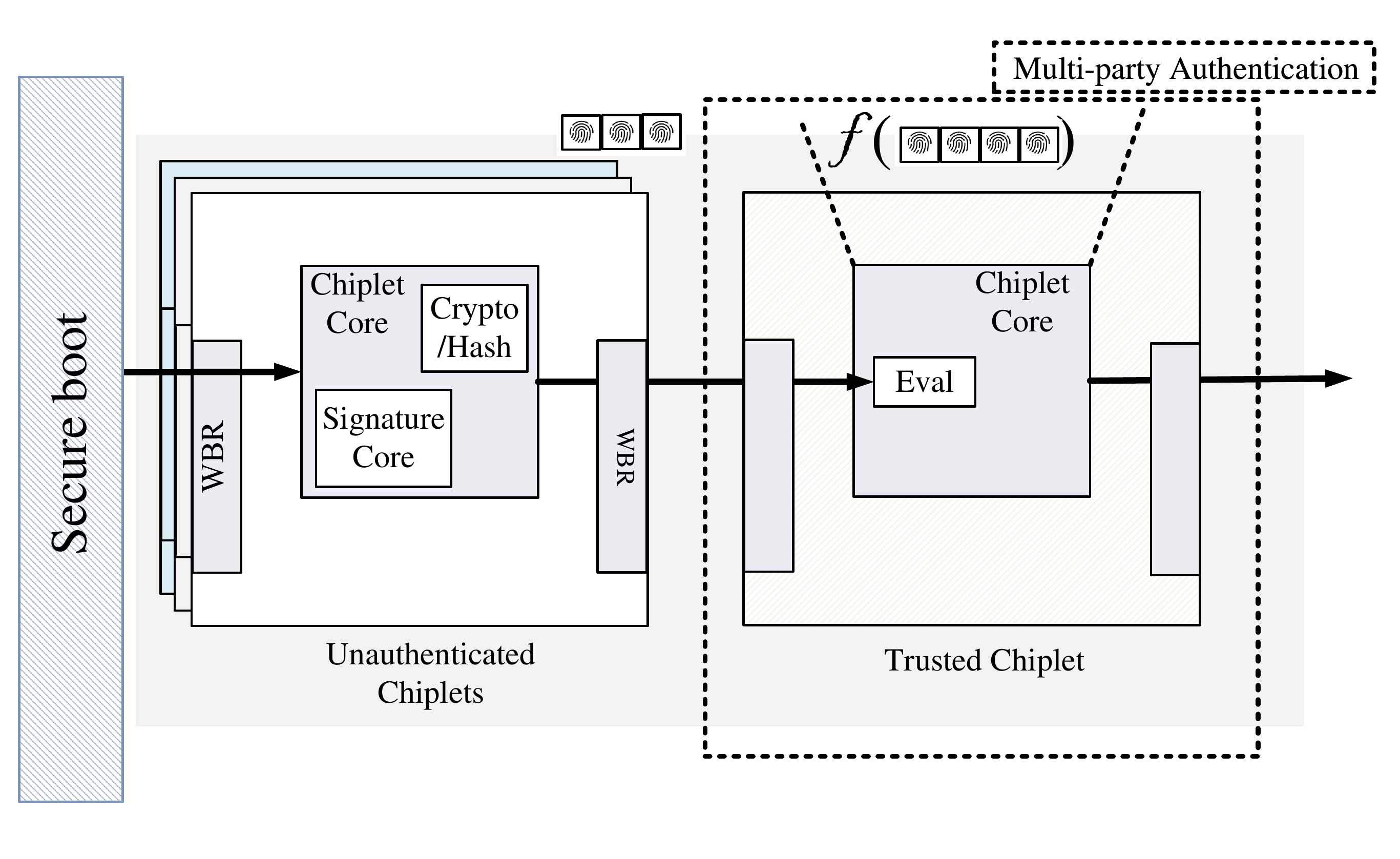}
    \caption{
        \emph{AuthenTree} Architecture. 
        Unauthenticated chiplets generate signatures and apply existing cryptographic/hash functions. Results are passed through standard Design-for-Test (DfT) interfaces, then validated by trusted chiplets using multi-party authentication.
    }
    \label{fig:authentree_arch}
\end{figure}

\section{Methodology}
\label{sec:method}

\noindent
This section presents a high-level overview and the architecture of the proposed \emph{AuthenTree} framework for distributed authentication in heterogeneous SiP systems.

\subsection{Overview and Scope of Work}
\label{subsec:overview}

\noindent
Guided by the core challenges discussed in Sec.~\ref{subsec:motivation}, \emph{AuthenTree} forgoes reliance on any single point of validation. Instead, it establishes a tree of trust—\circled{1}~where integrator chiplets initially cross-validate one another to build a trustworthy set, which then collaboratively authenticates the rest of the system. Authentication responsibilities are shared across this network, so that no single chiplet holds ultimate authority or exposes the system to a single point of failure. \circled{2}~This inherently supports scalability, as the architecture enables seamless addition or removal of chiplets and adapts flexibly to changes in system composition. Each authentication round exploits parallelism, such as distributed hashing and signature processing, across the tree, ensuring that strong security can be achieved efficiently. The final consensus check requires only a single comparator cycle. \circled{3}~Dynamic adaptability is achieved by partitioning all sensitive signature material and securely sharing it among integrators, such that no party can reconstruct or manipulate the complete authentication result alone. Notably, this approach is agnostic to chiplet orientation or placement strategies, relying solely on interposer-level interconnects. As a result, \emph{AuthenTree} remains inherently adaptable and can be deployed across diverse SiP topologies and integration styles.

It is important to note that \emph{AuthenTree}’s authentication protocol is designed to ensure the authenticity and integrity of chiplets, but does not explicitly detect or identify functional Trojans within chiplets themselves; addressing such concerns requires complementary secure testing methodologies, which are out of scope for this work. Similarly, the framework assumes PUF-based signatures or other hard-to-clone primitives, mitigating risks of signature copying by adversaries. Potential interposer-level attacks are addressed via cross-checking and protocol redundancy, making it difficult for any manipulation at the interconnect or packaging level to evade detection during validation.

The following sections detail the construction of this architecture, its protocol flow, and practical considerations for scalable, secure deployment in diverse SiP environments.




\begin{figure}[t]
    \centering
    \includegraphics[width=\linewidth]{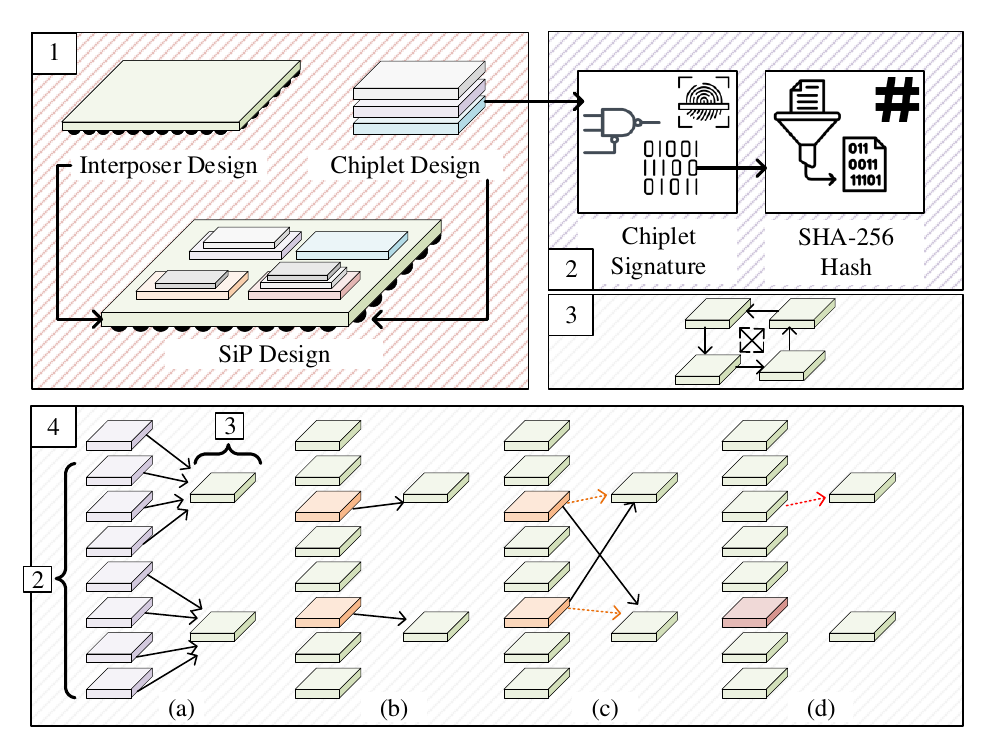}
    \caption{
        Flow of the \emph{AuthenTree} authentication process.
        \textbf{(1)}~SiP, interposer, and chiplet co-design. 
        \textbf{(2)}~Chiplet signatures are hashed using \emph{AuthenTree}, mitigating the risk of adversarial replication or eavesdropping.
        \textbf{(3)}~Integrator chiplets establish mutual trust via cross-authentication.
        \textbf{(4)}~\emph{Authentication workflow:}
        \textbf{(a)}~All third-party chiplets undergo authentication; 
        \textbf{(b)}~two chiplets return inconsistent results, indicating a possible chiplet or interconnect fault; 
        \textbf{(c)}~cross-validation is performed with additional trusted integrators; 
        \textbf{(d)}~final resolution finds interconnect issue and unauthenticated chiplet. 
    }
    \label{fig:authentree_flow}
\end{figure}

\subsection{Architecture}
\label{subsec:architecture}

\noindent
The architecture \emph{AuthenTree} builds on the observation that modern chips typically already incorporate strong and unreplicable signature cores, such as those derived from PUFs or scan-chain-based primitives~\cite{zheng2013scanpuf}, as well as standard cryptographic or hash cores for secure operation. In cases where a dedicated hash engine is absent, a lightweight SHA-256 implementation can be seamlessly integrated with minimal overhead~\cite{tashdid2025safe}. Unlike SAFE-SiP, which is based on centralized garbling and hash units, \emph{AuthenTree} eliminates these centralized elements and distributes authentication using a tree-based hierarchical protocol between integrator chips. This distributed approach naturally maps to the native SiP interconnect structure, requiring no modification to standard plug-and-play chiplet routing~\cite{li2024lucie} and imposing virtually no additional area or wiring overhead. The resulting system enables robust, privacy-preserving authentication that scales to diverse SiP topologies and supports dynamic chiplet addition or removal. Fig.~\ref{fig:authentree_arch} depicts the overall architecture, highlighting the integration of authentication within secure boot and Design-for-Test~(DfT) infrastructure.

\subsection{Proposed Flow}
\label{subsec:proposed_flow}

\noindent
Figure~\ref{fig:authentree_flow} outlines the complete operational flow of the \emph{AuthenTree} framework, from initial design through distributed authentication and fault localization in heterogeneous SiP systems.

\subsubsection{SiP Assembly and Signature Setup}
\noindent
The flow begins with SiP assembly at the foundry, where chiplets sourced from multiple third-party vendors are integrated onto a common interposer. Since these chiplets arrive as black-box components, it is essential to verify their authenticity after SiP packaging. Each chiplet is provisioned with a unique signature, ideally based on a strong, unclonable primitive such as a scan-chain PUF~\cite{zheng2013scanpuf}. To safeguard this identity from adversarial observation or replay, the signature is hashed via a lightweight cryptographic core (e.g., SHA-256), producing an obfuscated but verifiable token for later authentication.

\subsubsection{Signature Obfuscation}

\noindent
Prior to authentication, each chiplet's unique signature is processed through a cryptographic hash function, converting it into a fixed-length digest regardless of the original signature size. This not only standardizes signature lengths for subsequent aggregation but also conceals the underlying signature, preventing adversaries from inferring or reconstructing sensitive chiplet information. The resulting digest serves as a privacy-preserving anchor for secure, distributed authentication in the following protocol steps.

\subsubsection{Trusted Cross-Authentication}

\noindent
A subset of integrator chiplets is first selected to establish a tree of trust. They perform mutual, cross-authentication to validate each other's integrity before proceeding. This ensures that no single integrator chiplet holds ultimate authority, reducing the risk of a single point of failure or collusion. Once consensus is reached among this trusted set, they coordinate the authentication of all remaining third-party or untrusted chiplets.

\subsubsection{Hierarchical Authentication and Fault Localization}

\noindent
The authentication protocol in \emph{AuthenTree} is organized as a hierarchical, multi-stage process that systematically verifies every chiplet within the SiP while enabling precise fault localization. First, in step (a), all untrusted or third-party chiplets are authenticated by a quorum of integrator chiplets—each of which has previously cross-validated with its peers to establish a trusted validation set. This ensures that no single integrator can compromise the authentication decision. When a chiplet fails authentication, either due to inconsistent hash output, missing participation, or mismatched session response, it is flagged as anomalous in step (b). Rather than immediately disqualifying the chiplet, the protocol initiates a second round of verification (step c): the suspected chiplet is authenticated again, this time using a different subset or path of integrator chiplets and, where possible, alternative interposer routes. This cross-validation is critical for distinguishing between true chiplet faults (e.g., counterfeit, cloned, or unresponsive devices) and system-level issues such as interposer connectivity failures, routing congestion, or transient noise. By triangulating the results from multiple independent integrators and routing paths, \emph{AuthenTree} can reliably infer whether the authentication failure is attributable to the chiplet itself or to the surrounding interconnect infrastructure. Finally, as depicted in step (d), the framework uses these redundant, overlapping checks to localize faults: if multiple independent routes flag the same chiplet, it is conclusively identified as unauthenticated and is rejected from integration. Conversely, if authentication only fails along specific routes or with certain integrators, the fault is localized to an interposer or connectivity issue, prompting targeted remediation rather than chiplet exclusion. This hierarchical and distributed process minimizes false positives, improves system resilience, and maintains robust security even as the scale and complexity of the SiP increases.

\section{Security Analysis}
\label{sec:security-analysis}

\noindent
\emph{AuthenTree} is engineered to deliver reliable authentication and strong defense against a wide spectrum of attacks throughout the SiP supply chain. By decentralizing trust and employing a tree-based MPC protocol, the framework enhances system integrity and withstands adversarial conditions. Below, we discuss key attack vectors and analyze how \emph{AuthenTree} addresses each.

\begin{figure}[t]
    \centering
    \includegraphics[width=\linewidth]{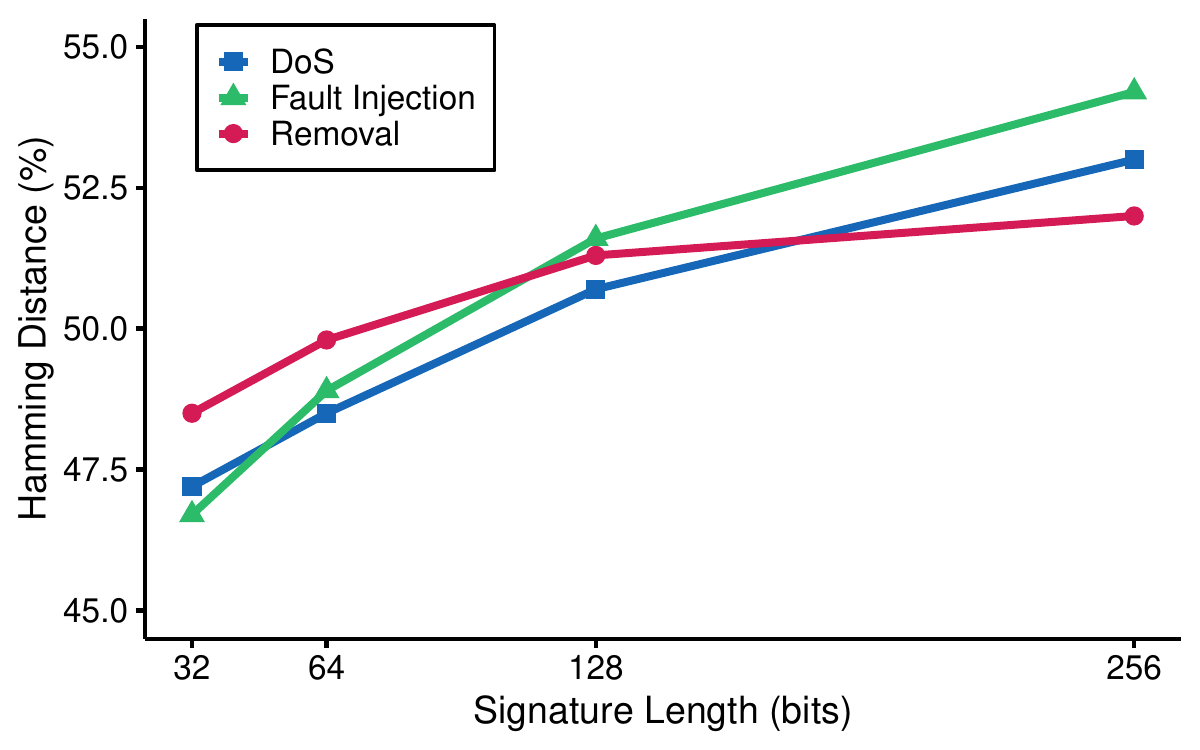}
    \caption{Hamming distance under removal, DoS, and fault injection attacks for varying signature lengths.}
    \label{fig:hamming_attacklines}
\end{figure}

\subsection{Compromise and Forgery}

\noindent
A core vulnerability in heterogeneous SiP platforms is the risk of chiplet compromise---either through insertion of malicious components, substitution of counterfeit devices, or modification of authentic modules. \emph{AuthenTree} defends against such threats by mandating that successful authentication requires consensus from a threshold number of integrator chiplets, each entrusted with only a portion of the overall secret via secure sharing. As no single integrator possesses enough information to reconstruct or manipulate the complete signature, adversaries must compromise a significant subset of the trusted nodes to succeed. This threshold-based approach substantially raises the difficulty of both compromise and forgery, effectively preventing unauthorized parties from producing valid authentication responses. All signatures are processed in a distributed fashion, ensuring both confidentiality and integrity throughout the supply chain.

\begin{table}[t]
\centering
\setlength\tabcolsep{1pt}
\caption{Design Area Overhead of \emph{AuthenTree}.}
\label{tab:authentree_design_area}
\begin{tabular}{lccc}
\toprule
\textbf{Design} & \textbf{Design Area} (\(\mu m^2\)) & \textbf{\emph{AuthenTree} Area} (\(\mu m^2\)) & \textbf{Overhead} (\%) \\
\midrule
CVA6~\cite{cva6}     &   345,755  & 6,988.20  & 2.02 \\
NVDLA~\cite{nvdla_hw}    &   541,552  & 7,192.12  & 1.33 \\
RISC-V~\cite{riscv_soc}   & 1,309,680  & 7,063.03  & 0.54 \\
Ariane~\cite{ariane}   & 1,431,536  & 7,000.50  & 0.49 \\
OR1200~\cite{ARM}   & 1,488,384  & 7,108.00  & 0.48 \\
\bottomrule
\end{tabular}
\end{table}

\subsection{Tampering and Fault Injection}

\noindent
The integrity of authentication can be threatened by active tampering or fault injection attacks---including deliberate glitches, voltage fluctuations, or targeted bit-level modifications. To rigorously assess \emph{AuthenTree}'s resilience, we systematically flip one or more bits in the input signature and measure the resulting Hamming distance between the SHA-256 hash of the original and tampered input. As shown in Fig.~\ref{fig:hamming_attacklines}, even minimal changes to the input generate large, unpredictable changes in the output hash, owing to the strong avalanche property of SHA-256. This ensures that any tampering, no matter how subtle, is reliably detected as a failed authentication. Such sensitivity is essential for real-world deployment, where hardware faults or adversarial interventions can manifest as only minor deviations in input data.

\subsection{Denial-of-Service (DoS)}

\noindent
Distributed systems are also susceptible to denial-of-service attacks, where one or more parties attempt to disrupt operation by withholding or corrupting their authentication shares. We model this threat by omitting an integrator’s contribution during the authentication process and analyzing the impact on the collective authentication output. Experimental results indicate that removal of any single share causes the output hash to differ substantially from the expected value, as reflected by a high Hamming distance across all tested signature lengths (see Fig.~\ref{fig:hamming_attacklines}). This property guarantees that incomplete or non-participatory authentication attempts are easily identified, making it infeasible for an adversary to subvert the protocol through selective denial or non-cooperation.

\subsection{Replay and Replication}

\noindent
Replay and transcript replication attacks occur when adversaries attempt to reuse valid authentication responses in subsequent sessions, hoping to impersonate legitimate chiplets. \emph{AuthenTree} neutralizes this risk by embedding a fresh, unpredictable nonce into every authentication session. The combination of unique session randomness and the collision resistance of SHA-256 ensures that each authentication instance is cryptographically bound to its context. This design makes it computationally infeasible to find different inputs producing the same hash output or to successfully replay prior authentication messages. The use of session-bound hashing thus prevents both accidental and malicious replay, reinforcing overall system security.

In sum, \emph{AuthenTree} delivers comprehensive security against a spectrum of threats---from compromise and tampering to denial-of-service and replay---leveraging decentralized trust and robust cryptographic design to ensure resilient and scalable SiP authentication.

\section{Design Overhead Analysis}
\label{sec:results}

\noindent
We evaluate \emph{AuthenTree} using a combination of RTL simulation and ASIC synthesis to demonstrate its scalability, efficiency, and security in heterogeneous SiP environments. We report authentication latency, area, and power, and provide detailed comparisons against prior approaches serving as baselines ~\cite{gate_sip,pqc_hi, sect_hi,tashdid2025safe}.

\textit{
While the main goal of this work is secure authentication for chiplet-based heterogeneous integration, security inevitably comes with costs, namely, increased power, performance, and area (PPA) overhead. To be viable for real-world adoption, any solution must minimize these penalties. We therefore evaluate PPA overheads and directly compare our approach with previous solutions, demonstrating that other existing frameworks are more infeasible than \emph{AuthenTree}.
}

\subsection{Experimental Setup}

\noindent
\emph{AuthenTree} was implemented in Verilog and evaluated using both Synopsys Design Compiler (SAED $14\,nm$ library) and Xilinx Vivado for FPGA prototyping. Power, area, and timing results were extracted from post-synthesis reports with the clock set at $1\,\mathrm{GHz}$.

\begin{table}[t]
\centering
\setlength\tabcolsep{2pt}
\caption{Area Comparison with Recent Works.}
\label{tab:area_comparison_recent}
\begin{tabular}{lccc}
\toprule
\textbf{Work} & \textbf{LUT Resource} & \textbf{FF} & \textbf{Area (mm$^2$)} \\
\midrule
SECT-HI~\cite{sect_hi} & -- & -- & $5.11$ \\
PQC-HI (Kyber+Dilithium)~\cite{pqc_hi} & $76,999$ & $49,993$ & --  \\
PQC-HI (Kyber only)~\cite{pqc_hi} & $1,842$ & $1,634$ & -- \\
SAFE-SiP ($\kappa=64$)~\cite{tashdid2025safe} & -- & -- & $0.0996$ \\
\textbf{\emph{AuthenTree}} & $\textbf{1740}$ & $\textbf{1054}$ & $\textbf{0.0071}$ \\

\bottomrule
\end{tabular}
\end{table}

\subsection{Area Overhead Analysis}

\noindent
We evaluated the area impact of \emph{AuthenTree} across a range of representative designs synthesized using a $14\,\mathrm{nm}$ technology node. As detailed in Table~\ref{tab:authentree_design_area}, \emph{AuthenTree} introduces only minimal design area overhead, ranging from $0.48\%$ for OR1200 to $2.02\%$ for CVA6, with an absolute hardware footprint of $\sim7{,}000~\mu m^2$ consistently observed across all evaluated platforms. This level of overhead is well within practical limits for state-of-the-art heterogeneous SiP systems, ensuring that robust security can be integrated without compromising design efficiency or scalability. Table~\ref{tab:area_comparison_recent} further situates \emph{AuthenTree} in the context of prior works, revealing that contemporary solutions such as SECT-HI~\cite{sect_hi} and PQC-HI~\cite{pqc_hi} incur substantially higher resource requirements: for instance, SECT-HI reports a physical area of $5.11~\mathrm{mm}^2$, while PQC-HI demands up to $76,999$ LUTs. By contrast, \emph{AuthenTree} delivers the smallest hardware footprint, with just $1,740$ LUTs, $1,054$ flip-flops, and a physical area of only $0.0071~\mathrm{mm}^2$. This translates to an area reduction of nearly $\textbf{14}\times$ compared to SAFE-SiP~\cite{tashdid2025safe} and over $\textbf{700}\times$ relative to SECT-HI, highlighting the exceptional efficiency and scalability of our distributed architecture. It is also important to note that GATE-SiP~\cite{gate_sip} was not included in this comparison, as the authors did not provide any area estimates or resource breakdowns for their design.

\begin{table}[t]
\centering
\setlength\tabcolsep{8pt}
\caption{Authentication Latency Comparison}
\label{tab:latency_comparison}
\begin{tabular}{lcc}
\toprule
\textbf{Work} & \textbf{Reported Latency} & \textbf{Notes} \\
\midrule
SECT-HI~\cite{sect_hi} & $280~\mathrm{ms}$ & Serialized \\
PQC-HI~\cite{pqc_hi} & $1$--$10~\mathrm{ms}$ & PQC operations \\
SAFE-SiP~\cite{tashdid2025safe} & $\sim1~\mu\mathrm{s}$ & Parallelizable \\
\textbf{\emph{AuthenTree}} & $<1~\mu\mathrm{s}$ & One-time, parallelizable \\
\bottomrule
\end{tabular}
\end{table}

\subsection{Latency and Runtime Performance}

\begin{figure}[t]
    \centering
    \includegraphics[width=0.97\linewidth]{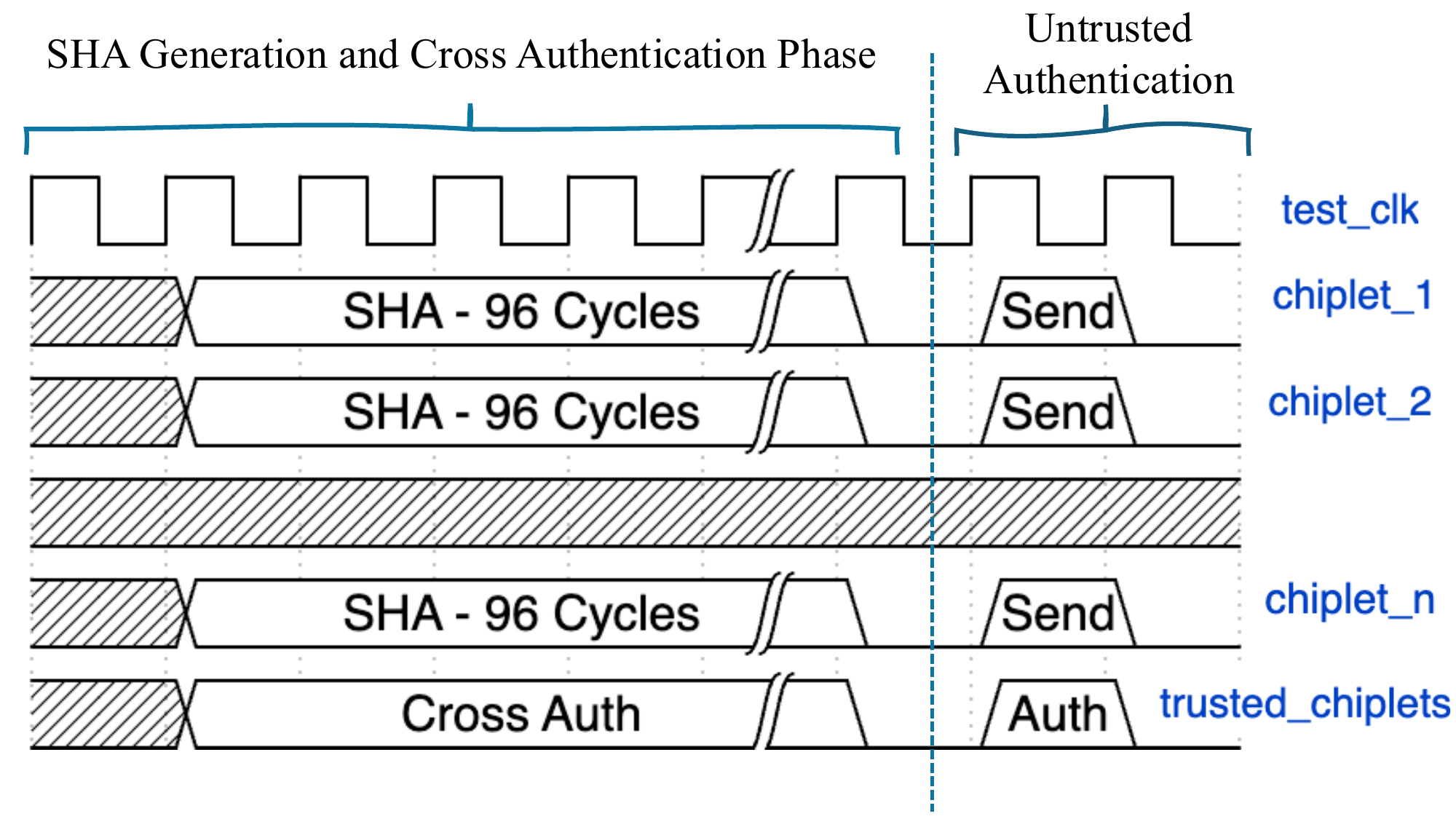}
    \caption{
    Timing diagram of the \emph{AuthenTree} protocol. All chiplets perform parallel SHA-256 signature generation and cross authentication within $96$ cycles, followed by a single round of result aggregation and authentication. The separation highlights trusted, parallelizable phases versus the untrusted authentication step.
    }
    \label{fig:authentree_timing}
\end{figure}

\noindent
Authentication latency is a critical factor for secure SiP systems, as high overhead can impede scalability and slow integration. In \emph{AuthenTree}, authentication is performed using a SHA-256 core that requires only $96$ cycles at $1\,\mathrm{GHz}$ (See Fig.~\ref{fig:authentree_timing}), resulting in a latency of less than $1~\mu\mathrm{s}$ per session. Importantly, this authentication is a one-time event for the system, and, owing to the distributed architecture, can be executed in parallel with chiplet integration or other operations. Furthermore, individual chiplets may leverage their own crypto cores to perform authentication locally, but any associated latency is still a one-time cost at setup and does not impact runtime operation. By comparison, prior works report far higher authentication delays: SECT-HI~\cite{sect_hi} incurs a $280~\mathrm{ms}$ delay, while PQC-HI~\cite{pqc_hi} suffers $1$--$10~\mathrm{ms}$ per round due to heavy cryptographic operations. SAFE-SiP~\cite{tashdid2025safe} achieves comparable sub-microsecond latency through parallelism, but its scalability is limited for large-scale distributed systems. Table~\ref{tab:latency_comparison} summarizes these results. Overall, \emph{AuthenTree} delivers a lightweight, one-time authentication protocol with minimal latency overhead, enabling scalable and efficient deployment in heterogeneous SiPs.

\subsection{Power Overhead Analysis}

\noindent
The dynamic and static power consumption of the authentication core is a key metric for large-scale and energy-efficient SiP deployments. As presented in Table~\ref{tab:authentree_power_overhead}, \emph{AuthenTree} achieves exceptionally low power overheads, ranging from $0.13\%$ (NVDLA) to $1.83\%$ (CVA6), with an absolute power increase of less than $0.26~\mathrm{mW}$ across all tested designs. This marks a substantial improvement over prior work such as \emph{SAFE-SiP}~\cite{tashdid2025safe}, where power overheads reached as high as $34\%$ due to the use of a full Yao-based garbling circuit~\cite{gc_yao}. In contrast, \emph{AuthenTree} replaces the costly garbling logic with a streamlined SHA-256 hashing core, an optimization enabled by our finding that hashing alone is sufficient for robust, non-malleable authentication. This design shift results in dramatically lower switching and internal power while preserving strong security guarantees.

Direct power comparisons with SECT-HI, PQC-HI, and GATE-SiP were not possible, as these works did not report quantitative power metrics. Nevertheless, their reliance on resource and arithmetic-intensive cryptographic primitives, such as block ciphers, authenticated encryption, or post-quantum algorithms, implies substantially greater power consumption than our hash-based approach.

\subsection{Discussion}

\noindent
This work systematically addresses the core challenges of secure, scalable authentication in chiplet-based SiP architectures. We explore and resolve each of our guiding research questions (RQs) through the design, implementation, and evaluation of the \emph{AuthenTree} framework.

\uline{\textbf{(RQ1)}}
One of the most pressing needs in heterogeneous SiP security is to avoid placing trust in any single chiplet or centralized anchor. \emph{AuthenTree} directly accomplishes this by decentralizing authentication using a tree-based Multi-party Computation (MPC) approach, in which multiple independent integrator chiplets must cooperate to validate system integrity. No single party possesses enough information or authority to unilaterally authenticate or compromise the system. This distribution of trust not only hardens the SiP against insider threats and targeted attacks but also aligns with zero-trust principles, removing any inherent reliance on proprietary or opaque security blocks.

\begin{table}[t]
\centering
\setlength\tabcolsep{6pt}
\setlength\extrarowheight{2pt}
\caption{Power Overhead Analysis for \emph{AuthenTree}.}
\label{tab:authentree_power_overhead}
\begin{tabular}{lccc}
\toprule
\textbf{Design} & 
\textbf{\begin{tabular}[c]{@{}c@{}}Baseline Power \\ (mW)\end{tabular}} & 
\textbf{\begin{tabular}[c]{@{}c@{}}\emph{AuthenTree} Power \\ (mW)\end{tabular}} & 
\textbf{\begin{tabular}[c]{@{}c@{}}Overhead \\ (\%)\end{tabular}} \\
\midrule
CVA6~\cite{cva6}   & $12.896$  & $0.2357$ & $1.83$ \\
NVDLA~\cite{nvdla_hw}  & $185.140$ & $0.2394$ & $0.13$ \\
RISC-V~\cite{riscv_soc} & $59.164$  & $0.2518$ & $0.43$ \\
Ariane~\cite{ariane} & $94.157$  & $0.2463$ & $0.26$ \\
OR1200~\cite{ARM} & $106.610$ & $0.2485$ & $0.23$ \\
\bottomrule
\end{tabular}
\end{table}

\uline{\textbf{(RQ2)}}
As SiP systems grow in scale and complexity, it becomes crucial to detect and isolate malicious or malfunctioning participants without sacrificing performance. \emph{AuthenTree}'s design ensures that each integrator's contribution is cryptographically essential: even the omission or manipulation of a single share leads to a clear authentication failure. The SHA-256-based protocol amplifies any tampering, making it immediately apparent and rendering attacks such as forgery, bypass, or fault injection ineffective. Moreover, the lightweight computational and hardware requirements guarantee that security does not come at the cost of efficiency, allowing the system to maintain consistent performance even as the number of chiplets or integrators increases.

\uline{\textbf{(RQ3)}}
Modern SiP design environments demand flexible integration, where chiplets from different vendors can be dynamically added, replaced, or removed to meet changing requirements. \emph{AuthenTree} is inherently compatible with these dynamic workflows. The protocol allows the authentication tree and secret sharing structure to be easily reconfigured for new participants, without global resets or full system re-authorization. This property not only streamlines supply chain management but also ensures that authentication remains fresh and secure across all operating conditions, even as system composition changes over time. The result is a security architecture that grows with the platform, enabling truly modular, plug-and-play integration while maintaining uncompromising security.

Overall, our approach delivers on every front: it provides an excellent and in fact, better alternative to traditional centralized schemes, scales efficiently with system size, and adapts seamlessly to the evolving realities of heterogeneous, open-market chiplet integration. The \emph{AuthenTree} framework thus represents a decisive step forward in the practical realization of secure, distributed, and future-proof SiP authentication.

\section{Conclusion}
\label{sec:conclusion}

\noindent
In this work, we introduced \emph{AuthenTree}, a scalable, distributed authentication framework for heterogeneous SiP platforms that eliminates single points of failure and central trust anchors. By leveraging multi-party computation and a tree-based architecture, \emph{AuthenTree} achieves robust, plug-and-play security with negligible area, power, and latency overheads, while enabling dynamic integration and strong resilience against a wide spectrum of supply chain threats. Our extensive evaluation demonstrates that \emph{AuthenTree} meets the demanding requirements of modern chiplet ecosystems, providing a practical path toward secure, future-ready System-in-Package deployments. This framework not only addresses today’s security needs but lays the groundwork for agile, modular, and trustworthy semiconductor innovation in an increasingly open and collaborative landscape.

\newpage

\bibliography{sample-base}

\end{document}